\documentstyle[epsf,epsfig,aps]{revtex}
\draft
\input epsf
\begin{document}
\twocolumn[\hsize\textwidth\columnwidth\hsize\csname@twocolumnfalse%
\endcsname
\title{Destruction of the phase coherence by the magnetic field in the fluctuation
region of thin superconducting film.}
\author{ \ A. \ V. \ Nikulov, \ S. \ V. \ Dubonos \ and \ Y. \ I. \ Koval}

\address{Institute of Microelectronics Technology and High Purity Materials, Russian Academy of Sciences, 142432 Chernogolovka, Moscow District, RUSSIA}

\maketitle
\begin{abstract}
{Reduction of the gradient of current density is observed near $T_{c}$ in
thin superconducting films. This effect is explained by the influence of
fluctuation superconductivity on the current density distribution. The
observed suppression of this effect by low magnetic field is explained by
the destruction of phase coherence by magnetic field. The current
density distribution in high magnetic field does not differ from that in
the normal state down to low temperature ($T < T_{c2}$). This is
interpreted as evidence that the superconducting state without pinning
below $H_{c2}$ is not a vortex liquid but a state without the phase
coherence. }
\end{abstract}
]
\narrowtext

	The superconductivity is a macroscopic quantum effect. Phase $\phi$ of
wave function $\Psi = |\Psi|exp(i\phi )$ of superconducting electrons can
cohere all over volume of a superconductor.
Quantization of a magnetic flux \cite{london} and the Meissner effect
\cite{meissner} are consequences of the phase coherence. According to the
relation for supercurrent (see for example \cite{huebener})
$$j = \frac{hen_{s}}{4\pi m}(\frac{d\phi}{dr} -
\frac{4\pi e}{hc}A) =\frac{1}{\lambda_{L}^{2}}(\frac{\Phi_{0}}{2\pi}
\frac{d\phi}{dr} - A) \eqno{(1)}$$
the magnetic flux, $\Phi$, contained within a closed path of integration
must be equal $\Phi = \Phi_{0} n$ because the integral of $d\phi /dr$ along
the closed path must be equal $n2\pi$. Here $\lambda_{L} =
(mc/e^{2}n_{s})^{0.5}$ is the London penetration depth; $n_{s}$ is the
superconducting electron density; $\Phi_{0} = hc/2e$ is the flux quanta; n
is an integer. n must be equal zero in the absence of singularity
(nonsuperconducting inclusion) within the closed path of
integration because in this case the closed path can be tightened to zero
while remaining inside superconducting region.
Thus, the magnetic flux cannot be within a superconducting region with
phase coherence and without singularities. And this is the cause of the
Meissner effect \cite{meissner}.

	Therefore, magnetic field can penetrate into a superconductor
only if: 1) superconductivity is destroyed, or 2) singularities
appears, or 3) the phase coherence is absent. The first case is observed in
type I superconductor. The second case is observed in the Abrikosov state
of type II superconductor. The singularities are the Abrikosov vorteces in
this case. These two cases are widely known. The third case is not so
well known. It is observed in fluctuation superconductivity
regions, for example above the second critical field, $H_{c2}$, in type II
superconductors.

	Zero electrical resistance, the main property of superconductors, can
exist only in states with a long range phase coherence. In a state
without the phase coherence superconducting
fluctuations increase the electrical conductivity \cite{nik81a} (this
excess conductivity is called paraconductivity) whereas resistance remains
non-zero.

	According to the Josephson relation \cite{joseph} electrical voltage
$$V = \frac{h}{4\pi e}\frac{d\phi}{dt} \eqno{(2)}$$
in states with the long range phase coherence. Here V is a voltage
between two points; $\phi$ is a phase difference between these points. The
phase difference changes in time in the Abrikosov state if the vorteces
flow. The resistance caused by the vortex flow is called flux flow
resistance \cite{huebener}. This name is no quite right because a magnetic
flux induced by a external magnet does not moved in a superconductor.

	The vortex flow is caused by an electrical current \cite{huebener}. A
flow velocity in real samples depends not only on a current value but also
on a pinning force \cite{huebener}. The vortex can not flow at a current
lesser than a critical current. The resistance is equal zero in this case.
The pinning effect may be in the state with the phase coherence only.
Therefore it may be considered as a consequence of the phase coherence.

	According to the Abrikosov solution \cite{abrikos} obtained in the
mean-field approximation the long range phase coherence appears at
mean-field transition point, $H_{c2}$. But the second critical field is no
a critical point in the fluctuation theory. In the consequence of the
fluctuation the pinning appears below $H_{c2}$ \cite{nik81a,nik84}. It was
shown in work \cite{nik81} that the pinning appearance in bulk
superconductor is a narrow transition. This transition was interpreted in
work \cite{nik81} as a transition from a fluctuation state, as the state
with superconducting electrons but without the phase coherence, into the
Abrikosov state, as the state with the long range phase coherence. The
position of this transition was denoted as $H_{c4}$ in work \cite{nik90}.

	The result of \cite{nik81} was repeated at investigations of
$YBa_{2}Cu_{3}O_{7-x}$ \cite{safar}. But the absence of the pinning in a
substantial range of temperature and magnetic field below $H_{c2}$ in
high-Tc superconductor was interpreted as a consequence of the vortex
lattice melting \cite{bishop}. This interpretation differs from the one
proposed in \cite{nik81}. The state without pinning is considered as a
vortex liquid in this interpretation. The vortex liquid is a state with the
phase coherence because the vorteces are singularities in superconducting
state with the phase coherence. And singularities can not exist without
medium.

	The resistance in the vortex liquid is caused by the vortex flow
\cite{bishop} as well as in the Abrikosov state. Therefore some difference
of resistive properties of the vortex liquid and of the state without the
phase coherence may be expected. Nonlocal resistivity in the vortex liquid
is predicted in paper \cite{huse}. The nonlocal resistivity is
considered as a consequence of the viscosity and the incompressibility of
the vortex liquid in \cite{huse}. The nonlocal resistivity in
the vortex liquid may be considered also as a consequence of the phase
coherence because the viscosity of the vortex liquid as well as the
vorteces can not exist without the phase coherence.

The nonlocal resistivity considered in \cite{huse} may be interpreted
as a reduction of a gradient of the current density in a consequence of the
viscosity of the vortex liquid. A similar nonlocal resistivity may be
observed also in the fluctuation region above $T_{c}$. According to the
relation (1) and the Maxwell equation a superconducting current density
within fluctuation superconducting drops which appear near $T_{c}$ must be
constant because size of this drops is smaller than the penetration depth.
Therefore a gradient of the current density must decrease near $T_{c}$.
This decreasing can be visible if size of the fluctuation superconducting
drops is no much small. It may be within the critical region. The critical
region of a bulk superconductor is very narrow. Therefore we use amorphous
$Nb_{1-x}O_{x}$ thin films. The critical region of these films is enough
wide.

The $Nb_{1-x}O_{x}$ films were produced by magnetron sputtering of Nb in
an atmosphere of argon and oxygen. The critical temperature $T_{c}$ of
these films depend on oxygen content. Three films with $x \simeq 0.15 -
0.25$ and $T_{c}$ 1.86 K, 2.46 K and 2.52 K were used in this work.
Amorphous $Nb_{1-x}O_{x}$ is a type-II superconductor with big value of the
Ginzburg-Landau parameter $\kappa$ and of the normal resistivity,
$\rho_{n}$: $\kappa \simeq 50$, $dH_{c2}/dT = -22\; kOe/K$, $\rho_{n}=200\;
\mu \Omega \ cm$. The temperature dependence of normal resistivity is very
weak, $|d\rho_{n}/dT|/\rho_{n} < 0.0002\; K^{-1}$ in the region 20-40 K,
where superconducting fluctuation is small. The resistivity increases with
decreasing temperature. This change can be connected with weak
localization. The used films have very weak pinning. It was shown in work
\cite{nik95} that the resistive dependencies of these films are described
by paraconductivity theory in a wide region both above and below $H_{c2}$.

	The film structure used is shown schematically in Fig. 1. This
structure was obtained by electron lithography and ion-beam etching. The
width of the strip in which current flows is equal approximately $4 \mu m$.
The branch width is equal approximately $8 \mu m$. The distance between
potential contacts 1-1 is equal approximately $6 \mu m$. The distance
between potential contacts 3 and 4 is $4 \mu m$, between 3 and 2 is $8
\mu m$. The width of the strip of the potential contacts is $0.2 \mu m$.
The film thickness d = 20 nm.

The voltage was measured with a relative error 0.0001. The temperature
was measured with a relative error 0.001. A magnetic field produced by a
superconducting solenoid was perpendicular to a film plane.

\begin{figure}[bhb] \vspace{0.1cm}\hspace{-1.5cm}
\vbox{\hfil\epsfig{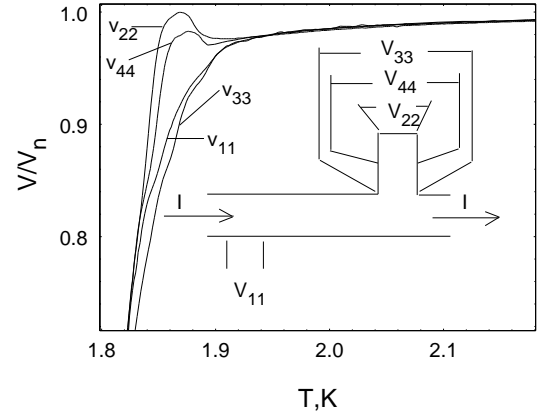}\hfil}
\vspace{0.75cm} \caption{Temperature dependencies of voltage relation, $V/V_{n}$, on
different potential contacts in the zero magnetic field. $I = 2 \ \mu A$. The
film structure and the location of the potential contacts are shown
schematically in the inset.} \label{fig-1} \end{figure}

Results obtained by investigation of three samples with $T_{c}$ = 1.86
K, 2.46 K and 2.52 K are similar. The temperature dependencies of the
relation $v_{ii} = V_{ii}(T)/V_{ii,n}$ on contacts 1-1, 4-4, 2-2 and 3-3 of
the sample with $T_{c}$ = 1.86 K in zero magnetic field are shown on figure
1. $V_{ii}(T)$ is voltage on contacts 1-1, 4-4, 2-2 and 3-3 at temperature
T and $V_{ii,n}$ is the same at T = 4.2 K where superconducting fluctuation
is very small. The relations voltage on the contacts to the transport
current value, I, in the normal state are $V_{11,n}/I = 146\; \Omega $;
$V_{33,n}/I = 185\; \Omega $; $V_{44,n}/I = 111 \; \Omega $; $V_{22,n}/I = 60 \;
\Omega $. The $v_{ii}(T)$ do not depend on the transport current value if I
does not heat appreciably the film. The overheating of the region between
contacts 1-1 determined by displacement of the resistive transition at
relatively large current is equal $0.008 \;K/(10 \ \mu A)^{2}$. The
overheating of the regions between other contacts is smaller. The
dependencies on Fig.1 are obtained at $I = 2 \ \mu A$. The overheating at
this current value does not exceed 0.0005 K.

Because the film is very homogeneous (critical temperature difference of
regions between contacts 2-2, 3-3, 4-4 and 1-1 does not exceed 0.001 K) the
resistivity, $\rho$, between all contacts must be equal. Consequently the
observed difference of the $v_{22}$, $v_{44}$, $v_{33}$ and $v_{11}$ values
is caused by a change of the current density, j, between contacts 2-2, 3-3,
4-4 near $T_{c}$. It is obvious that j between contacts 1-1 does not change
if I is constant. The $v_{11}(T,H)$ dependencies repeat the resistivity
dependencies obtained before \cite{nik95}. These dependencies are described
by paraconductivity theory \cite{nik95}. $1/v_{11} -1$ is proportional to
the paraconductivity value, $\sigma_{fl}$: $1/v_{11} -1 = \rho_{n}/\rho(T)
-1 = \sigma_{fl}\rho_{n} $. If we will introduce an average current density
between the potential contacts, $j_{ii}$, we may write $V_{ii} = j_{ii}\rho
l_{ii}$.  $l_{ii}$ is a distance between potential contacts i-i. Then
$v_{ii} = V_{ii}(T)/V_{ii,n} = j_{ii}(T)\rho (T)/j_{ii,n}\rho_{n}$. Where
$j_{ii,n}$ is the average current density between the contacts i-i in the
normal state without superconducting fluctuation (at T = 4.2 K). Because
$j_{11}$ is not changed, $v_{ii}/v_{11} = j_{ii}(T)/j_{ii,n}$. We may
estimate a value of a additional current between contacts i-i near the
superconducting transition from above relation. $\Delta j_{ii} = j_{ii}(T)
- j_{ii,n} = j_{ii,n}(v_{ii}/v_{11} - 1)$.

\begin{figure}[bhb] \vspace{0.1cm}\hspace{-1.5cm}
\vbox{\hfil\epsfig{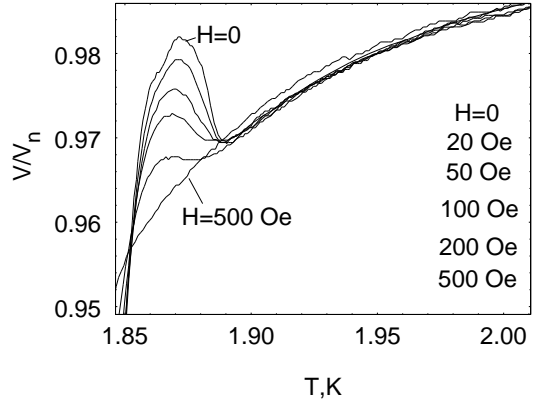}\hfil}
\vspace{0.75cm} \caption{Temperature dependencies of voltage relation on
potential contacts 4-4, $V_{44}/V_{44,n}$, at different values of
perpendicular magnetic field. The upper (at T = 1.87 K) curve - H = 0, the
lower (at T = 1.87 K) curve - H = 500 Oe.} \label{fig-1} \end{figure}

It is obvious that the difference of the values $V_{33,n}/I$;
$V_{44,n}/I$; $V_{22,n}/I$ is caused first of all by a difference of the
current density between these contacts, $j_{22,n} < j_{44,n} < j_{33,n}$,
because distances between these potential contacts are approximately equal.
Near $T_{c}$, $v_{44}/v_{11} - 1 > 0$, $v_{22}/v_{11} - 1 > 0$ and
$v_{33}/v_{11} - 1 < 0$ (Fig.1). This means that the current density
between the different contacts draw nearer close by $T_{c}$. According to
figure 1 $v_{22}/v_{11} - 1 > v_{44}/v_{11} - 1$. This means that the
additional current value decreases with removal from the I strip more
slowly than $j_{ii,n}$. $v_{44}/v_{11} - 1$ and $v_{22}/v_{11} - 1$
increases with approach to the $T_{c}$ more sharply than paraconductivity
value and is observed near $T_{c}=1.84 \;K$ only (Fig.1). The $T_{c}$ value
was determined from paraconductivity dependence in linear approximation
region with exactness 0.01 K.

Thus, the reduction of the gradient of the current density near $T_{c}$
is observed indeed. This effect was observed in all investigated samples
and therefore can not be connected with sample inhomogeneity. Following
to \cite{huse} we will call this effect as the nonlocal resistivity. The
nonlocal resistivity can be useful for investigation of an action of a
magnetic field on the superconducting drops. Two possibility can be in a
enough high magnetic field: 1) the Abrikosov vorteces appear within the
superconducting drops; 2) the magnetic field destroys the phase coherence.
In the first case the observed effect must not decrease in the magnetic
field because the relation (1) is linear.

\begin{figure}[bhb] \vspace{0.1cm}\hspace{-1.5cm}
\vbox{\hfil\epsfig{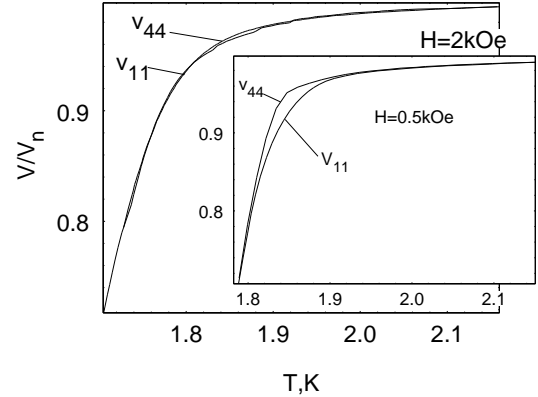}\hfil}
\vspace{0.75cm} \caption{Temperature dependencies of voltage relation on potential contacts
4-4, $V_{44}/V_{44,n}$, and 1-1, $V_{11}/V_{11,n}$, in the perpendicular
magnetic field H = 0.5 kOe and H = 2 kOe. $I = 2 \ \mu A$.} \label{fig-1} \end{figure}

	Our investigations show that a magnetic field suppresses the nonlocal
resistivity much more strongly than the paraconductivity value (Fig.2). The
nonlocal resistivity begins to decrease in very low magnetic field (Fig.2),
decreases in some times at H = 0.5 kOe and disappears in higher magnetic
field (fig.3). The paraconductivity value changes very little in these
values of the magnetic field (see Fig.2). The observed small change of the
paraconductivity value is consistent with small change of the
mean-field critical temperature, $T_{c2} - T_{c} = H/(dH_{c2}/dT)$.  At H =
20 Oe, $T_{c2} - T_{c} \simeq -0.001\; K$; at H = 0.5 kOe, $T_{c2} - T_{c}
\simeq -0.02 \;K$.

The magnetic field can not act on superconducting drops with size
smaller than $(\Phi_{0}/H)^{0.5}$. Consequently the superconducting drops
with size $ > (\Phi_{0}/20 \ Oe)^{0.5} = 1 \mu m$ contribute to the observed
effect. But the nonlocal resistivity is observed at H = 500 Oe. Therefore
we can propose that the superconducting drops with size $ < (\Phi_{0}/500 \
Oe)^{0.5} = 0.2 \mu m$ contribute to the effect also.

Thus, the magnetic field destroys the phase coherence in the fluctuation
superconducting drops first of all. The nonlocal resistivity predicted for
the vortex liquid in work \cite{huse} is not observed in high magnetic
field up to low temperature. This means that a state without pinning is no
the vortex liquid but is the state with superconducting electrons but
without the phase coherence. Consequently the transition observed in
\cite{nik81,safar} is not the vortex lattice melting but is the phase
coherence disappearance.

A.V.N. is grateful to M.A.Moore for the idea to measure the nonlocal
resistivity in a thin superconducting film. This work was supported by the
National Program of Russia on "Superconductivity" of SSTP "ADPCM", Project
95040.

\end{document}